\documentclass[aps,twocolumn,float,superscriptaddress,prd,nofootinbib]{revtex4-1}
\usepackage{graphicx, epsfig, bm, amsmath}
\usepackage{color}
\usepackage{hyperref}
\usepackage{ wasysym }
\usepackage{array}

\begin{document}

\newcommand{\lcdm}{$\Lambda$CDM}
\newcommand{\gpr}{G^{\prime}}
\newcommand{\fnl}{f_{\rm NL}}
\newcommand{\curv}{{\cal R}}
\newcommand{\R}{\mathcal{R}}
\newcommand{\dotR}{\dot{\mathcal{R}}}
\newcommand{\ddotR}{\ddot{\mathcal{R}}}
\newcommand{\ep}{\epsilon_H}
\newcommand{\dotep}{\dot{\epsilon}_H}
\newcommand{\et}{\eta_H}
\newcommand{\dotet}{\dot{\eta}_H}
\newcommand{\cs}{c_s}
\newcommand{\esq}{\left(}  
\newcommand{\dir}{\right)} 
\newcommand{\ord}{\mathcal{O}}
\newcommand{\hR}{\hat{\mathcal{R}}}
\newcommand{\dothR}{\dot{\hat{\mathcal{R}}}}
\newcommand{\mpl}{M_{\text{pl}}}

\definecolor{darkgreen}{cmyk}{0.85,0.2,1.00,0.2}
\newcommand{\vin}[1]{\textcolor{darkgreen}{[{\bf VM}: #1]}}
\newcommand{\wh}[1]{\textcolor{blue}{[{\bf WH}: #1]}}
\newcommand{\peter}[1]{\textcolor{red}{[{\bf PA}: #1]}}

\title{Steps to  Reconcile Inflationary Tensor and Scalar Spectra}

\author{  Vin\'icius Miranda }
\affiliation{Department of Astronomy \& Astrophysics, University of Chicago, Chicago IL 60637}
\affiliation{The Capes Foundation, Ministry of Education of Brazil, Bras\'ilia DF 70359-970, Brazil}
\author{ Wayne Hu}
\affiliation{Department of Astronomy \& Astrophysics, University of Chicago, Chicago IL 60637}
\affiliation{Kavli Institute for Cosmological Physics,  Enrico Fermi Institute, University of Chicago, Chicago, IL 60637}
\author{ Peter Adshead}
\affiliation{Department of Physics, University of Illinois at Urbana-Champaign, Urbana, IL 61801, U.S.A.}

\begin{abstract}
The recent BICEP2 B-mode polarization determination of an inflationary tensor-scalar
ratio $r=0.2^{+0.07}_{-0.05}$ is in tension with simple scale-free models of inflation due to
a lack of a corresponding low multipole excess in the temperature power spectrum which places
a limit of $r_{0.002}<0.11$ (95\% CL) on such models.   Single-field inflationary models that reconcile these two observations, even those where the tilt runs substantially, introduce a scale into the scalar power spectrum.   To cancel the tensor excess, and simultaneously explain the excess already present in $\Lambda$CDM, ideally the model should introduce this scale as a relatively 
sharp transition in the tensor-scalar ratio around the horizon at recombination.   We consider
models which generate such a step in this quantity and find that they can improve the
joint fit to the temperature and polarization data by up to $2\Delta \ln{\cal L} \approx -14$ without changing cosmological parameters.  Precision E-mode polarization measurements
should be able to test this explanation.
\end{abstract}

\maketitle

\section{Introduction}

The recent BICEP2 measurement of a tensor-scalar ratio $r=0.2^{+0.07}_{-0.05}$ from degree scale B-mode polarization of the cosmic-microwave background (CMB) \cite{Ade:2014xna} is in ``moderately-strong" tension with slow-roll inflation models that predict scale-free, albeit slightly tilted ($1-n_s  \ll 1$) power-law  power spectra. This tension is due to the implied excess in 
 the temperature spectrum at low multipoles which is not observed and restricts $r_{0.002}< 0.11$ (95\% CL) in this context  \cite{Ade:2013uln}.   
 
 These findings can be reconciled in the single-field inflationary paradigm by introducing a scale
into the scalar power spectra to suppress power on these large-angular scales.  For example a large running of tilt, $dn_s/d\ln k \sim -0.02$, is possible as a compromise \cite{Ade:2014xna}.
Here the scale introduced is associated with the scalar spectrum transiently passing through
a scale-invariant slope near observed scales.
 However, such a large running is uncomfortable in the simplest models of inflation which typically produce running of order $\mathcal{O}[\left(1-n_s\right)^2]$. Moreover, a large running also requires further additional parameters in order that inflation does not end too quickly after the observed scales leave the horizon \cite{Easther:2006tv}.

The temperature anisotropy excess implied by tensors is also not a smooth function of scale,
but rather cut off at the horizon at recombination.    To counter this excess, 
a transition in the scalar power spectrum that occurs more sharply, though coincidentally near these scales, would be preferred.
Such a transition can occur without affecting the tensor spectrum if there is a slow-roll
violating step in the tensor-scalar ratio while the Hubble rate is left nearly fixed.   
In this work we consider the effects of placing such a feature near scales associated with the horizon at recombination, thereby suppressing the scalar spectrum on large scales. 

This slow-roll violating behavior also produces oscillations in the power spectrum \cite{Adams:2001vc, Peiris:2003ff, Park:2012rh, Miranda:2012rm} and generates enhanced non-Gaussianity \cite{Chen:2006xjb, Chen:2008wn} if this transition occurs in much less than an efold.  For transitions that alleviate the tensor-scalar tension, these oscillations would violate tight constraints on the acoustic peaks and hence only transitions that occur over at least
an efold are allowed.   The resulting non-Gaussianity is then undetectable \cite{Adshead:2011bw, Adshead:2012xz}.
Throughout, we work in natural units where the reduced Planck mass $M_{\rm Pl} = (8\pi G_N)^{-1/2} = 1$ as well as $c = \hbar =  1$.

\section{Step Solutions}
 \label{sec:step}
 
\begin{figure}[t]  
\psfig{file=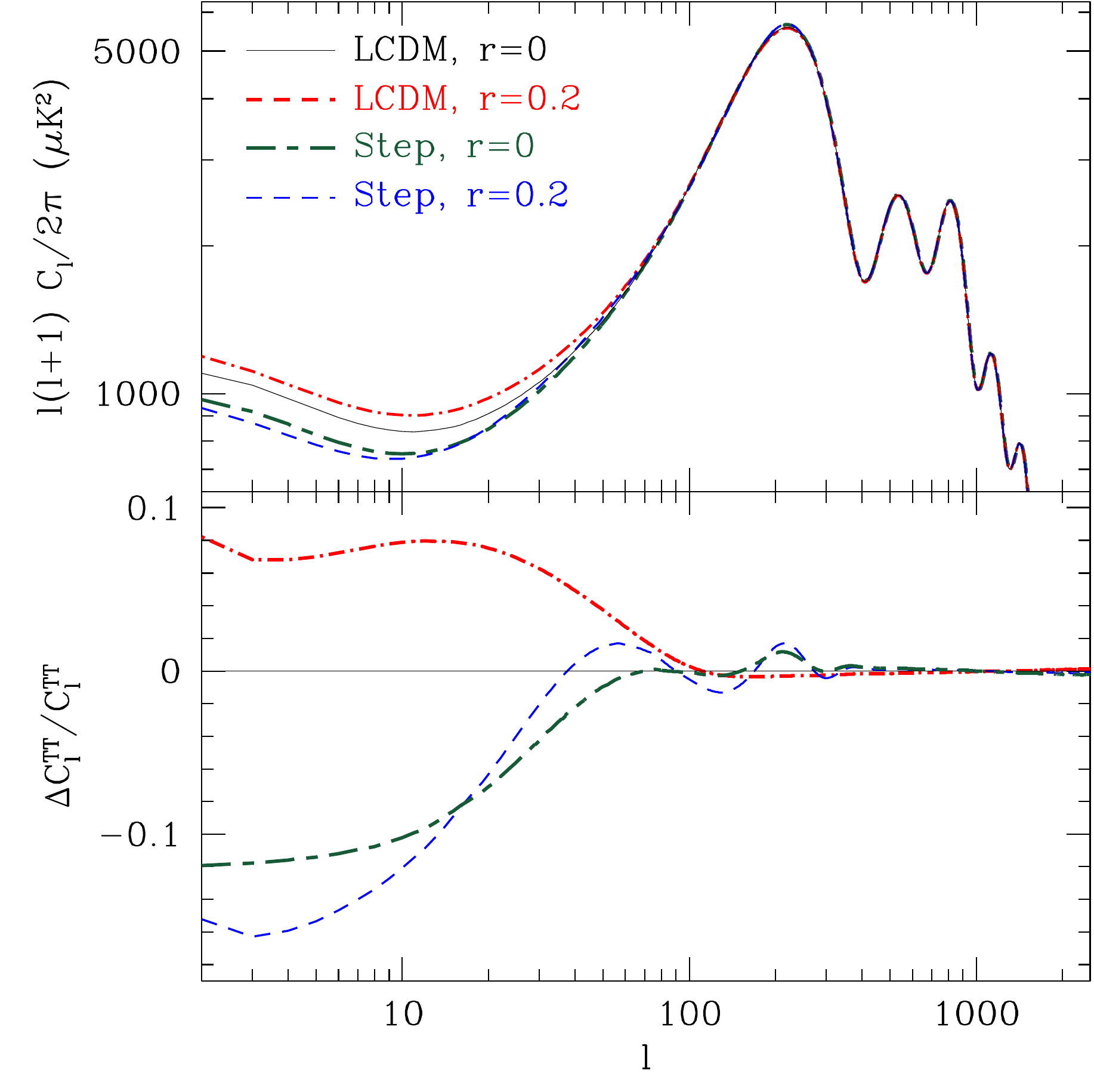, width=3.25in}
\caption{Total temperature power spectra showing the unobserved excess produced by 
adding tensors of $r=0.2$ to the best fit 6 parameter $\Lambda$CDM model and its
removal by adding a step in the tensor-scalar parameter  $\epsilon_H c_s$.   Planck data in fact favor removing more power than the tensor excess, preferring a step even if $r=0$.  Step model parameters are given in Tab.~\ref{tab:fits}. }
\label{plot:cl}	
\end{figure}

 In slow roll inflation, the tensor power spectrum in each gravitational wave polarization state is directly related to the Hubble scale during inflation 
  \begin{equation}
 \Delta_{+,\times}^2 = \frac{H^2}{2\pi^2},
 \end{equation}
 whereas the scalar or curvature power spectrum is given by
 \begin{equation}
 \Delta_\curv^2 = \frac{H^2 }{8\pi^2 \epsilon_H c_s},
 \end{equation}
 where $\epsilon_H = -d\ln H/d\ln a$ and $c_s$ is the sound speed, yielding a
 tensor-scalar ratio $r= 4 \Delta_{+,\times}^2 /  \Delta_\curv^2 = 16\epsilon_H c_s$.  
 The addition of a nearly scale invariant tensor spectrum to the CMB temperature 
 anisotropy produces excess power below $\ell\approx 100$ which at $r=0.2$ is difficult
 to accommodate in slow roll inflation where the scalar spectrum is, to a good 
 approximation, a scale-free power law (see Fig.~\ref{plot:cl}).      
 
 The scalar power spectrum can be changed largely without affecting the tensors if
 the quantity $\epsilon_H c_s$ changes while $\epsilon_H$ remains small.   
 As shown in Fig.~\ref{plot:cl}, the excess power
 resembles a step in this quantity on scales near the horizon at recombination.
 Hence to alleviate the tension between the tensor inference from the BICEP2 experiment,
 $r=0.2^{+0.07}_{-0.05}$, 
 and the upper limits from the combined CMB temperature power spectrum $r_{0.002}< 0.11$ (95\% CL), we examine
 models where there is a step in this quantity.
 In this paper we quote $r$ at the scalar pivot
 of $k=0.05$ Mpc$^{-1}$ where it is unaffected by changes to the scalar power spectrum that
 we introduce whereas the upper limit is quoted at $k=0.002$ Mpc$^{-1}$.

 As an example, we consider a step in the
 warp 
\begin{align}\label{eqn:T(phi)}
	T(\phi) =& \frac{\phi^4}{\lambda_B} \left\{1 + b_T \left[ \tanh\Big(\frac{\phi-\phi_s}{d}\Big)-1 \right] \right\}
\end{align}	
 of Dirac-Born-Infeld (DBI) inflation\footnote{Of course, we are well outside the region of validity of UV complete versions of DBI inflation. However, this is merely a phenomenological proof of principle rather than a working construction.} \cite{Silverstein:2003hf, Alishahiha:2004eh} with the Lagrangian
 \begin{equation}
{\cal L}= \left[
1-\sqrt{1 - 2  X/T(\phi)} \right] T(\phi)- V(\phi),
\label{eqn:DBI}
\end{equation}
where the kinetic term  $2X = - \nabla^{\mu} \phi \nabla_{\mu} \phi$, 
  the sound speed
\begin{equation}
c_s(\phi,X) =\sqrt{ 1 - 2 X/T(\phi)}.
\end{equation}
  Here $\{ b_T,\phi_s,d \}$ parameterize the 
height, field position and field width of the step while the underlying parameters
$\lambda_B$ and the inflaton potential $V(\phi)$ are set  to
 to fix $n_s$ and $A_s$ \cite{Adshead:2013zfa}.   In Ref.~\cite{Miranda:2012rm},
    we showed that such a model produces a step in the quantity $\epsilon_H c_s$ that controls the tensor-scalar ratio.
To keep this discussion model independent, we follow Ref.~\cite{Miranda:2013wxa} and quantify the amplitude of the step by 
the change in this quantity
\begin{equation}
C_1 = -\ln \frac{ \epsilon_{Hb} c_{sb}}{ \epsilon_{Ha} c_{sa}},
\end{equation}
where ``$b$" and ``$a$" denote the quantities before and after the step on the slow roll 
attractor.   For definiteness, we take $c_{sb}\approx 1$.
In place of $\phi_s$ we quote the sound horizon 
\begin{equation}
s = \int d{N} \frac{c_s }{a H}
\end{equation}
 at the step $s_s=s(\phi_s)$ and in place
of the width in field space $d$, we take the inverse of the number of efolds $N$ the inflaton
takes in traversing the step
\begin{equation}
x_d = \frac{1}{\pi d} \frac{ d\phi}{d\ln s}.
\end{equation}   See Ref.\ \cite{Adshead:2011jq, Miranda:2013wxa} for details of this description.  
We utilize the generalized slow roll technique \cite{Stewart:2001cd, Dvorkin:2009ne, Hu:2011vr} to calculate the power spectra
of these models since at the step the slow roll approximation is transiently violated. 

 \begin{figure}[t]  
\psfig{file=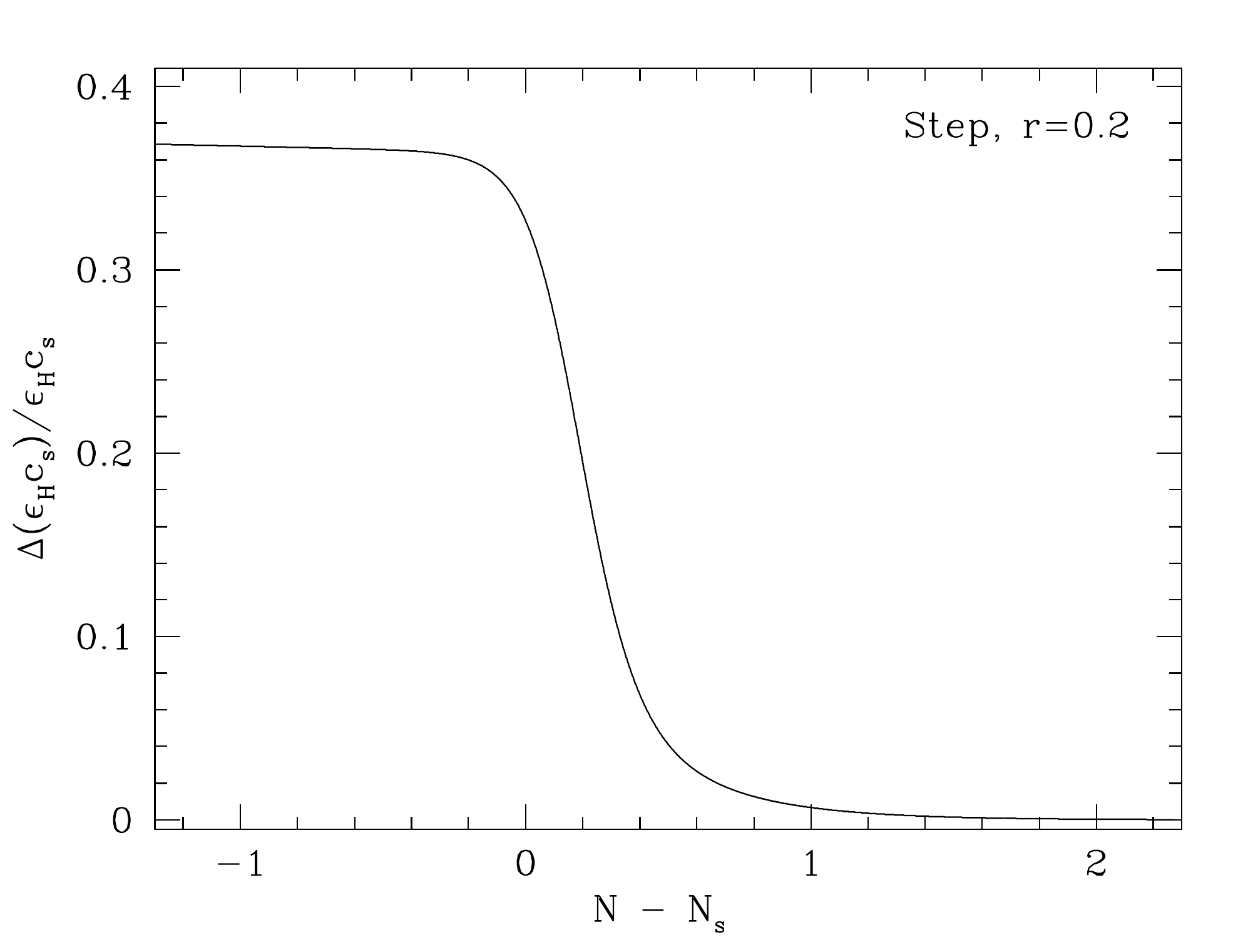, width=3.25in}
\caption{Step in tensor-scalar ratio parameter $\epsilon_H c_s$ relative to no step, from the best fit $r=0.2$ solution centered at the efold
$N_s$ at which the inflaton crosses the step.  Planck data favor a step that is traversed in about an efold.} \label{plot:epscs}	
\end{figure}

 \begin{figure}[t]  
\psfig{file=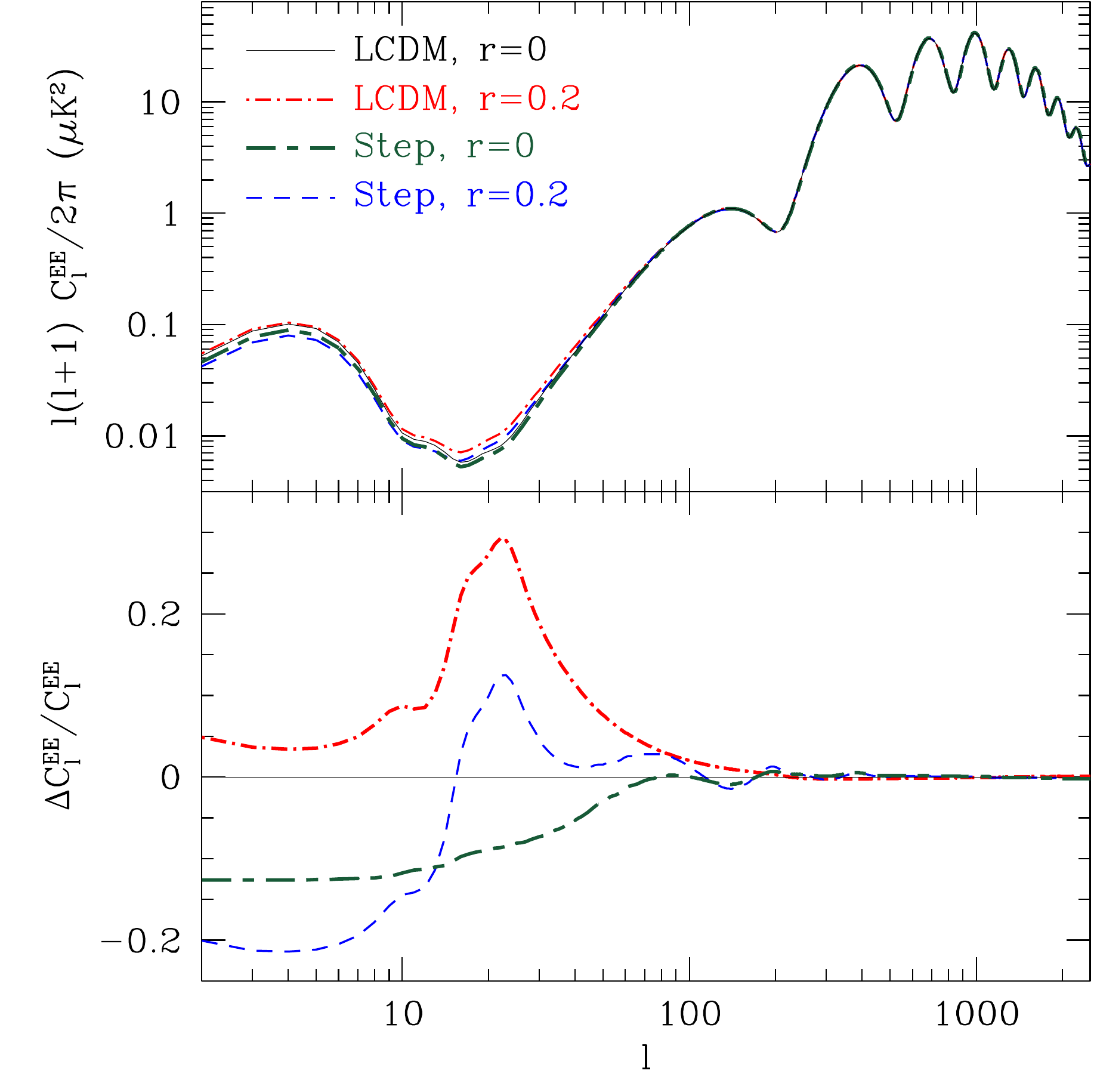, width=3.25in}
\caption{$EE$ power spectrum for the models in Fig.~\ref{plot:cl} showing the change from the
best fit $r=0$ $\Lambda$CDM power spectrum.    Excess $E$-modes from the tensors at $r=0.2$
are partially compensated by the step at $\ell \gtrsim 30$ while changes at lower $\ell$ can be altered
by changing the reionization history.  Preference for removing power at substantially smaller $r$ would predict a deficit of power as the $r=0$ model shows.} \label{plot:ee}	
\end{figure}

\begin{table*}[t] \centering
\begin{tabular}{ c | c | c| c| c|  c|  c|  c| c }
$r$ & $C_1$ & $s_s$(Mpc) & $x_d$ & $A_s \times 10^9$ &$n_s$& $-2\ln{\cal L}_P$ & $-2\ln {\cal L}_B $ & $-2\ln \Delta{\cal L}_{\rm tot}$ \\
\hline
0 & 0 & - & - & 2.1972 &  0.961 & 9802.7 & 89.1 & 40.1 \\
0 & -0.15  & 337.1 &  1.58 & 2.2003 &  0.957 &  9798.6 & 89.2  & 36.1 \\ 
0.1 & 0 & - & - & 2.1961 &  0.962 & 9806.5 & 47.9  & 2.7  \\ 
0.1 & -0.22 &  339.2 & 1.60 & 2.2000 &  0.958 &  9797.8 &  48.2 & -5.7 \\  
0.2 & 0 & - & -& 2.1939 &  0.963 & 9812.3 &  39.4 & 0 \\  
0.2 & -0.31 &  351.8 & 1.47 & 2.2002 &  0.959 &  9798.1 &  39.9 & -13.7 \\  
\hline  
\end{tabular}
\caption {Likelihood for models with tensors and steps with non-inflationary parameters fixed.  ${\cal L}_P$ is the likelihood for the Planck low-$\ell$ spectrum, high-$\ell$ spectrum and WMAP9 polarization; ${\cal L}_B$ is that for the BICEP2 $E$ and $B$ likelihood.  The change in the total is quoted relative to the  $r=0.2$ no feature case. }  
 \label{tab:fits} 
\end{table*}

\section{Joint Fit}

We jointly fit the Planck CMB temperature results, WMAP9 polarization results, and
BICEP2  to models with and without steps in the tensor-scalar ratio parameter $\epsilon_H c_s$.  We use the MIGRAD variable metric algorithm  from the CERN Minuit2 code \cite{James:1975dr} and a modified version of CAMB \cite{Lewis:1999bs,Howlett:2012mh} for model comparisons. The Planck likelihood includes the Planck low-$\ell$ spectrum (Commander, $\ell < 50$) and the high-$\ell$ spectrum (CAMspec, $50 <\ell < 2500$), whereas the BICEP2 likelihood\footnote{\href{http://bicepkeck.org/}{http://bicepkeck.org/}} includes both its $E$ and
$B$ contributions.

 We begin with the baseline best fit 
 6 parameter slow-roll flat $\Lambda$CDM model with $r=0$.   This model sets the
 non-inflationary cosmological parameters to
  $\Omega_c h^2= 0.1200$, $\Omega_b h^2=0.02204$, $h=0.672$, $\tau=0.0895$
  and the inflationary scalar amplitude  at $k=0.05$Mpc$^{-1}$, $A_s=2.1972 \times 10^{-9}$, and spectral tilt, $n_s=0.961$.    When considering alternate models we
 fix the non-inflationary parameters to these values while allowing the inflationary
 parameters, including $A_s$ and $n_s$ to vary.

  As shown in Tab.~\ref{tab:fits}, this $r=0$ model is strongly penalized
 by the BICEP2 data.   Moving to the $r=0.2$ model with the same parameters removes
 this penalty at the expense of making the Planck likelihood worse by $2\Delta\ln {\cal L}=
 9.6$ due to the excess in the $\ell \lesssim 100$ temperature power spectrum shown in
 Fig.~\ref{plot:cl}.

Next we fit for a step with parameters $C_1$, $s_s$, $x_d$ controlling the
amplitude, location and width of the step.   The best fit model at $r=0.2$ more than removes
the penalty from the temperature excess for Planck while fitting the BICEP2 $BB$ results
equally well.    The net result is a preference for a step feature at the level of $
2\Delta\ln {\cal L}_P=-14.2$ over no feature. The inclusion of BICEP2 results slightly degrades the fit to $
2\Delta\ln {\cal L}_{\rm tot}=-13.7$ due to changes in the $EE$ spectrum (see below).  
The $r=0.2$ model with a step is very close to the global maximum 
with further optimization in $r$ allowing only an improvement of $2\Delta\ln {\cal L}_{\rm tot}=-0.1$.
With the addition of the step, there remains a small high-$\ell$ change in the vicinity of the
first acoustic peak  in Fig.~\ref{plot:cl} which is interestingly marginally favored by 
the data.
Note that we have fixed the non-inflationary  parameters to their values without the step, for example $\tau$.
 Thus the likelihood 
may in fact increase in a full fit  (see Fig.~\ref{plot:ee}).  Conversely, we do not consider any compromise solutions where non-inflationary cosmological parameters ameliorate the tension without a step.  We leave these considerations to a future work.

The best fit step also predicts changes to the $EE$ polarization.  Like the $TT$ spectrum,
the excess power from the tensor contribution is partially  compensated by the reduction
in the scalar spectrum for $\ell \gtrsim 30$.  This is a  signature of the step model
which requires only a moderate increase in data to test as witnessed by the change in the
BICEP2 likelihood of $2\Delta \ln {\cal L}_B \sim 0.5$ it induces.  
Differences at $\ell \lesssim 30$, shown here at fixed $\tau$,
are largely degenerate with changes in the ionization history \cite{Mortonson:2009qv}

Due to potential contributions from foregrounds in the BICEP2 data
which may imply a shift to $r=0.16^{+0.06}_{-0.05}$ \cite{Ade:2014xna}, we also
test models at $r=0.1$ which would formally be in tension with the BICEP2 likelihood without
foreground subtraction.   Even in this case, the Planck portion of the likelihood improves
with the inclusion of a step though the preference is weakened to $2\Delta\ln{\cal L}_P=-8.6$
versus no step.   At $r=0$, the Planck data still prefers a step to remove power
at a reduced improvement of $2\Delta\ln{\cal L}_P=-4.1$, a 
fact that was already evident in the Planck collaboration analysis of anticorrelated isocurvature perturbations \cite{Ade:2013rta}.   Such an explanation should also help 
resolve the tensor-scalar tension albeit outside of the context of single-field inflation.
Interestingly, the addition of tensors at both $r=0.1$ and $0.2$ in fact further helps step models fit the Planck data
due to the changes shown in Fig.~\ref{plot:cl} independent of the BICEP2 result.

\section{Discussion}

A transient violation of slow-roll which generates a step in the scalar power spectrum at scales near to the horizon size at recombination can alleviate problems of predicted excess power in the temperature spectrum, present already in the best fit $\Lambda$CDM spectrum, and greatly
exacerbated by tensor contributions implied by the BICEP2 measurement.  
Such a step may be generated by a sharp change in the speed of the rolling of the inflaton $\epsilon_H$ or by a sharp change in the speed of sound $c_s$ over a period of around an efolding which combine to form the tensor-scalar ratio.   Preference for a step from the temperature power spectrum is at a level
of $2\Delta\ln {\cal L}_P = -14.2$  if $r=0.2$  and is still $-8.6$ at $r=0.1$, the lowest plausible
value that would fit the BICEP2 data.    

Such an explanation makes several concrete predictions.
Since slow-roll is transiently violated in this scenario, there will be an enhancement in the associated three-point correlation function. However, we do not expect this signal to be observable as it impacts only a small number of modes \cite{Adshead:2011bw, Adshead:2012xz}.  $E$-mode fluctuations on similar scales would be predicted to have
a smaller enhancement then with tensors alone.   This prediction should soon be testable; in the BICEP2 data it brings down the total likelihood improvement to 
$2\Delta\ln {\cal L}_{\rm tot}=-13.7$ with a step at $r=0.2$.

While we have used a DBI type Lagrangian to illustrate the impact of a change in the
tensor-scalar ratio parameter $\epsilon_H c_s$ due to a step in the sound speed, we do not expect that our results require this form, though precise details of the fit may change. Transient shifts in the speed of sound have been found to occur in inflationary models where additional heavy degrees of freedom have been integrated out \cite{Achucarro:2012yr}. We leave investigation of specific constructions to future work.

\acknowledgements

While this work was in preparation, the work \cite{Contaldi:2014zua} appeared which has some overlap with the work presented here. We thank Maur\'icio  Calv\~ao, Cora Dvorkin, Dan Grin, Chris Sheehy and Ioav Waga for useful conversations. This work was supported in part by the Kavli Institute for Cosmological Physics at the University of Chicago through grants NSF PHY-1125897 and an endowment from the Kavli Foundation and its founder Fred Kavli. WH was additionally supported by U.S.~Dept.\ of Energy contract DE-FG02-13ER41958 and VM by the Brazilian Research Agency
CAPES Foundation and by U.S. Fulbright Organization.

\vfill 
\break
\bibliography{bistep}
\vfill

\end{document}